\documentclass[conference]{IEEEtran}
\IEEEoverridecommandlockouts
\usepackage{cite}
\usepackage{amsmath,amssymb,amsfonts}
\usepackage{algorithmic}
\usepackage{graphicx}
\usepackage{multicol}
\usepackage[normalem]{ulem}
\usepackage[font={footnotesize}]{caption}
\usepackage{subcaption}
\usepackage[letterpaper, left=.91in, right=.91in, bottom=.9in, top=0.7in]{geometry}
\usepackage{textcomp}
\usepackage{xcolor}
\def\BibTeX{{\rm B\kern-.05em{\sc i\kern-.025em b}\kern-.08em
   T\kern-.1667em\lower.7ex\hbox{E}\kern-.125emX}}
\begin{document}

\title{Joint Functional Splitting and Content Placement for Green Hybrid CRAN
\thanks{This study is supported by EU Celtic-Plus Project SooGREEN: Service
Oriented Optimization of Green Mobile Networks.}
}

\author{\IEEEauthorblockN{Ajay Sriram\IEEEauthorrefmark{2}, Meysam~Masoudi\IEEEauthorrefmark{2}, Abdulrahman~Alabbasi\IEEEauthorrefmark{1}, and Cicek~Cavdar\IEEEauthorrefmark{2}}   \IEEEauthorblockA{\IEEEauthorrefmark{2}Communication Systems Department, KTH Royal Institute of Technology, \IEEEauthorrefmark{1}Ericsson Research, Stockholm, Sweden}
    Email: \IEEEauthorrefmark{2} \{ajays, masoudi, cavdar\}@kth.se, \IEEEauthorrefmark{1} abdulrahman.alabbasi@ericsson.com}
    

\maketitle
\begin{abstract}
A hybrid cloud radio access network (H-CRAN) architecture has been proposed to alleviate the midhaul capacity limitation in C-RAN. In this architecture, functional splitting is utilized to distribute the processing functions between a central cloud and edge clouds. The flexibility of selecting specific split point enables the H-CRAN designer to reduce midhaul bandwidth, or reduce latency, or save energy, or distribute the computation task depending on equipment availability. Meanwhile, techniques for caching are proposed to reduce content delivery latency and the required bandwidth. However, caching imposes new constraints on functional splitting. In this study, considering H-CRAN, a constraint programming problem is formulated to minimize the overall power consumption by selecting the optimal functional split point and content placement, taking into account the content access delay constraint. We also investigate the trade-off between the overall power consumption and occupied midhaul bandwidth in the network. Our results demonstrate that functional splitting together with enabling caching at edge clouds reduces not only content access delays but also fronthaul bandwidth consumption but at the expense of higher power consumption.
\end{abstract}


\section{Introduction}
Future mobile networks are required to support 1000-fold more traffic with the same cost and energy consumption. On the other hand, mobile services have a wide range of delay requirements \cite{zander2017beyond}. To meet such demands while supporting cost-effective network scalability, a cloud radio access network (C-RAN) architecture has been proposed to reduce energy consumption and network cost \cite{peng2014heterogeneous}. 
The main idea of C-RAN is to centralize all distributed digital units (DUs) along with their associated cooling into a centralized \emph{DU pool} at a central cloud (CC), only leaving behind the radio units (RUs) at the different cell sites. Such a centralization allows for reducing the energy consumption of RAN thanks to the network function virtualization and multiplexing gains. With C-RAN,  DUs can be shared among lightly loaded cells, which in turn enables load-dependent DU (de)activation/sleeping mechanisms. Accordingly, this can reduce network operational costs also spent on cooling and site maintenance, among a multitude of other benefits \cite{checko2015cloud}. 

Despite its appealing features, additional network challenges are introduced by C-RAN \cite{masoudi2017green}. Specifically, since fully-processed high-bandwidth radio-over-fiber (RoF) signals are now sent from the DU pool to the intended RUs, the fronthaul capacity between the CC and RUs becomes a transmission bottleneck, especially as the number and data rates of users served by the associated RU increase.  As per the recent literature, functional splitting \cite{virtualization2016functional} and caching content placement \cite{tran2017collaborative} are two potential solutions for this bottleneck problem. 


Functional splitting offers  additional freedom in dividing the signal processing between the CC and cell sites with several possible split points, provided that a cell site server is made available \cite{virtualization2016functional}. The survey in \cite{larsen2018survey} provides a detailed description of each functional split option and their advantages and disadvantages.
In \cite{maeder2014towards}, the  application and constraints of flexible C-RAN have been analyzed. In this study, the authors introduced a novel RAN-as-a-service concept, which leverages cloud technologies to implement a flexible functional split in 5G mobile networks. 
The authors in \cite{rost2014cloud}, discussed an architectural evolution from 3GPP LTE, outlined challenges and potential technologies to implement functional split, and described the potential gains. By flexible functional split in \cite{harutyunyan2017flexible}, the authors dealt with  decision making on the appropriate  functional split. The problem is formulated as an integer linear programming (ILP) problem whose objective is to jointly minimize the inter-cell interference and the fronthaul bandwidth utilization. In \cite{alabbasi2017delay}, an end-to-end delay model has been proposed for different functional split options. Considering this delay model, an optimization problem is formulated to reduce the system's power consumption and fronthaul bandwidth consumption.

Caching is  another approach that may not only relieve the fronthaul congestion but also can reduce the content delivery latency. In caching, popular contents are cached into a place closer to the users, e.g., in the edge cloud (EC), allowing user content demands to be accommodated more easily and quickly. Since content access delay is an important factor in caching problems, various algorithms and techniques have been proposed to incur lower latencies. The challenges, paradigm, and potential solutions for caching are discussed in \cite{tran2017collaborative}. In \cite{tran2016octopus}, cooperative hierarchical caching has been proposed to minimize the content access delay and boost the quality-of-experience (QoE) for end users. In \cite{kwak2018hybrid}, the authors proposed caching algorithms to optimize the content caching locations and hence reduce the delivery delay. In \cite{7881651} the authors presented a caching structure and proposed a cooperative multicast-aware caching strategy to reduce the average latency of delivering content. In this study, the focus is more on the users' QoE and the content access latency. 

Hybrid CRAN (H-CRAN) is proposed in \cite{alabbasi2017delay} to alleviate the limitations due to delay and fronthaul capacity. 
H-CRAN leverages the previous CC/EC structure with functional splitting in a three-layer architecture to share the processing tasks between CC and EC \cite{alabbasi2018optimal}. 
H-CRAN can simultaneously employ caching and functional splitting to tackle the fronthaul bottleneck problem. Although the existing content caching algorithms can reduce the service delay, it is not easy to decide where to deploy the content caches since there is a trade-off in balancing the centralized function processing and the distributed caching especially in H-CRAN. It is worth noting that caching the content  at the EC prevents us from functional splitting since the content is already at EC and it is not meaningful to centralize processing at CC. Due to this dependency, it is important to jointly decide whether to centralize or distribute content caching together with network processing functions. 
This paper is a first attempt to find a compromise between two contradictory trends: (1) Centralizing resources in radio access networks in C-RAN for energy and cost savings vs (2) Distributing the caching capabilities closer to the users in content delivery networks. With this regard, with the objective to minimize the network power consumption, we formulate the joint problem of content placement and flexible functional split in H-CRAN networks  as a \emph{constraint programing} \cite{manual2015ibm}. The problem formulation captures a delay constraint where each content should be delivered to the user within a delay threshold. Dependency of edge caching, content placement, and functional split decisions between edge and central cloud is formulated as a constraint. Moreover, we model the delay and power consumption network components from cloud to the user. Finally the trade off between the total system power consumption and bandwidth has been investigated. 

The rest of paper is organized as follows. In Section~\ref{Sec:system}, the network and reference architectures are explained. In Section~\ref{Sec:Prob}, the problem formulation is presented. In Section~\ref{Sec:Sim}, the network performance is evaluated.  Finally, the concluding remarks are discussed in Section~\ref{Sec:con}. 
\section{System Model}\label{Sec:system}
\subsection{Network Architecture}

\begin{figure}[!ht]
  \begin{center}
  \includegraphics[width=.8\columnwidth]{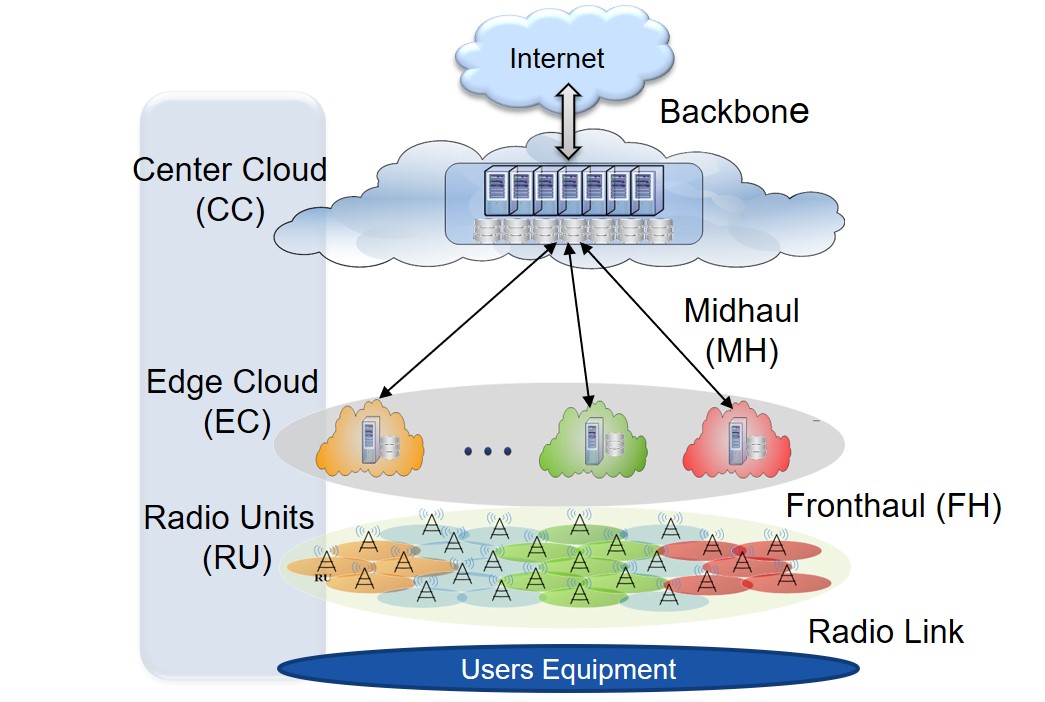}
  \end{center}
  \caption{Hybrid virtualized RAN architecture}
  \label{fig:H-CRAN}
\end{figure}

An H-CRAN architecture with dual-site processing is depicted in Fig. \ref{fig:H-CRAN}, where DUs are deployed at both the CC and the ECs. The three-layer architecture comprises a cell layer, an EC layer, and a CC layer. The term ``cell" is used here to refer to the coverage region of an RU. The cell layer consists of all cells, each serving several user equipment (UEs). Each group of cells are connected to an edge cloud as an aggregation point. The fronthaul between the CC and ECs can be implemented using a short fiber or wireless links, e.g., mm-Wave links or optical links.

 In this study, we consider mm-Wave because deployment of massive number of fiber links to support a densified cell layer can be costly. The ECs are connected to the CC via midhaul using various technologies, from expensive dark fiber solutions, to cost-efficient passive optical network (PON) families or other Ethernet based technologies. The midhaul technology considered in this study is time-wavelength division multiplexing PON (TWDM-PON) to allow time sharing of wavelengths between RUs, and each midhaul link is a wavelength channel, needing an optical network unit (ONU) at the EC and a line card (LC) at the CC as transceivers. The CC and ECs are equipped with DUs which enable a \emph{virtualized} functional processing, in which their computational resources can be virtualized and shared by connected RUs. For instance, the traffic to a cell can be partially processed at the CC so that the midhaul bandwidth requirement can be relaxed, then the remaining processing could be done at the EC. However, the processing at the EC is less efficient since the number of its accommodated DUs is less than that at the CC. Besides, the EC caching capacity is also less than that of the CC. The flexibility of the EC/CC functional processing also depends on the placement of the requested files. Hence, sharing infrastructure equipment based on the optimal cache placement creates a trade-off between the consumed power, midhaul bandwidth and experienced delay. Our quest hence becomes (1) whether to fully perform our baseband processing at the CC, the EC, or select an intermediate split point and (2) whether to place the content at the edge or central cloud given a user request with a delay threshold.

\subsection{Reference Architectures}
In this study, we refer to the link between the CC and the EC as the x-haul link. If all the processing is done at the EC then x-haul becomes backhaul. And if all the functions are centralized,  the corresponding x-haul link is called fronthaul. In case we partially centralize and split network functions between the CC and the EC, the x-haul link is called midhaul. The following two, fully-centralized and fully-distributed  reference architectures with no functional splitting are used as a baseline for performance evaluation purposes. 
\begin{enumerate}
\item The first reference case is when all the requested files are placed in EC, then all the baseband processing must be placed at the EC, and the connection from EC and CC is provided by backhaul. Since all the processing takes place at the EC, more power is consumed but in return we require less bandwidth in the backhaul. This is the best case in terms of satisfying users' delay requirements. 
\item The second reference case is when all the requested files are placed at the CC together with full centralization of network functions, introducing extra delay and consuming the highest x-haul bandwidth, with the advantage of minimized power consumption. Note that in this reference case the delay constraint needs to be relaxed. 
\end{enumerate}

\subsection{Functional Split model}
\begin{figure}[t]
  \begin{center}
   \includegraphics[width=.75\columnwidth]{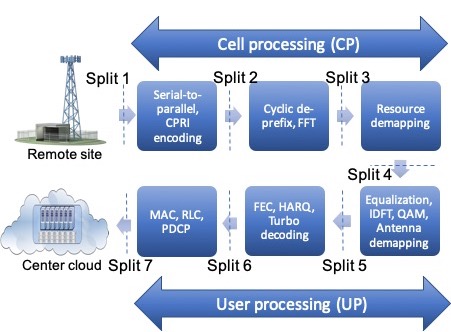}

  \end{center}
  \caption{Functional split model}
  \label{fig:FSmodel}
\end{figure}

The communication baseband processing contains a set of functions that can be classified as cell processing (CP) and user processing (UP) functions. In this paper, we represent the functional split of the baseband processing as shown in Fig. \ref{fig:FSmodel}. The CP mainly represents the functions within the physical layer that are responsible for  the signal processing associated with the cell. Few examples of CP functions include: CPRI encoding, cyclic (de)prefix, and resource (de)mapping. Similarly, the UP includes a set of functions that are related to physical layer and some upper layer functions that are responsible for signal processing of each user in a cell. Few examples of UP functions include: antenna (de)mapping, forward error correction (FEC). According to the Fig. \ref{fig:FSmodel}, the functional split can either happen before split 1 or after split 7 or in between. When split happens at Split 1, then all the functions are centralized at CC resulting in CRAN. When split happens after Split 7, all the functions are centralized at EC resulting in DRAN. When Split happens in between, the function above the split are placed at CC and function below the split are placed at EC.

\section{Problem Formulation and Solution Methodology} ~\label{Sec:Prob}
In this section, the problem of content placement and functional split in hybrid C-RAN is formulated as a Constraint programming problem as follows.

\subsection{Objective}
The objective is to minimize the network's power consumption. The objective function is expressed as 
\begin{equation}
min (P_t)
\end{equation}
where $P_t$ denotes the total power consumption in the network and is expressed as,
\begin{eqnarray}\nonumber
P_t&=&(g\times P_{LC}) \\\nonumber &+&\Big[\Big(P_{CC}^{cool} + lP_{CC}^{DU}\Big)I(l>0) +  P_{CC}^{cache}\Big] \\\nonumber&+&  \sum_{r\in \mathbb{R}}\Big[\Big(\sum_{c\in \mathbb{C}_r}(P_{Tx}+P_{FH})I(|\mathbb{I}_c| >0)\Big)\\\nonumber&+&\Big(I\Big(\sum_{c\in \mathbb{C}_r}|\mathbb{I}_c| >0\Big)P_{ONU} + P_{EC}^{cool}I(l_r>0)\\ &+& l_rP_{EC}^{DU} + I\Big(\sum_{c\in \mathbb{C}_r}\sum_{i\in \mathbb{I}_c}\delta_i>0\Big)P_{EC}^{cache}\Big)\Big]
\end{eqnarray}
The first term in (7), is the power consumption of line cards in the midhaul where $g$ is a sum over each wavelength calculating whether it is used by any EC. The second term is the power consumption of CC while the third term is that of each EC where $l$ and $l_r$ can be calculated similar to $g$.  $I$ is the indicator function which becomes 1 if the defined statement within brackets is true else 0. All terms of the power model are explained in Section \ref{sec:power}.

\subsection{Given}
\subsubsection{System Parameters}

\begin{itemize}
\item \textbf{Topology:} One CC is connected  to  multiple EC. Each EC is  connected  to  an  exclusive  set  of  cells,  and  each  cell exclusively covers a set of UEs.
\item \textbf{$\mathbb{I}_x$:} Set of UEs. When $x$ = 0, it refers to all UEs in H-CRAN, otherwise, it refers to set of UEs in cell c.
\item \textbf{$\mathbb{C}_x$:} Set of cells. When $x$ = 0, it refers to all cells in the entire H-CRAN, $x$ = $r$ refers to the set of cells belonging to EC $r$.
\item \textbf{$\mathbb{D}_x$:} Set of DUs. When $x$ = 0, it refers to all DUs in H-CRAN, $x$ = $-1$ refers to set of DUs in the CC, $x$ = $r$ refers to the set of DUs in EC $r$.
\item \textbf{$\mathbb{R}$:} Set of ECs.
\item \textbf{$\mathbb{W}$:} Set of wavelengths.
\item \textbf{$\mathbb{F}$:} Set of files to be cached.
\item \textbf{$F_x$:} Set of functional split options, where $x$ = $\{UP,CP\}$.
\item \textbf{$d_{i}$:} Delay threshold of a UE $i$.

\item \textbf{$H_x^y(.)$:} Pre-calculated mapping from a split option $x$ = $\{UP,CP\}$ to the number of (UP and CP) functions at site $y$ = $\{CC,EC\}$.
\item \textbf{$J_i(.)$:} Pre-calculated mapping from UP split of UE $i$ to the required midhaul bandwidth, which is proportional to the number of resource blocks (RBs) allocated to UE $i$.
\item \textbf{$G_c(.)$:} Pre-calculated mapping from CP split of cell $c$ to the required midhaul bandwidth, which is proportional to the number of antennas and carrier bandwidth.
\item \textbf{$K$:} Bandwidth capacity of a wavelength. Note that this is different from bandwidth induced and consumed by user’s and cell’s processing split, described with $J_i(.)$ and $G_c(.)$.
\item \textbf{$L_x^y$:} The capacity of a DU located at the ``$y$" site, $y$ = $\{CC,EC\}$, in terms of the number of $x$ functions that can be accommodated by this DU ($x$ represents CP or UP, and $y$ represents CS or RS). \textbf{Note: } $L_{CP}^{EC} < L_{CP}^{CC}$
\item \textbf{{$S_f$}:} Size of the content.
\item \textbf{{[$f_i,d_i$]}:} User demand pair which shows user i's request of content f within delay threshold $d_i$.
\item \textbf{$C_x$:} Maximum storage capacity of the cache at $x$ = $\{EC,CC\}$.
\end{itemize}

\subsubsection{Power calculation parameters}\label{sec:power}
\begin{itemize}

\item $P_{CC}^{DU}$, $P_{EC}^{DU}$: Power consumption of DU at CC/EC.
\item $P_{LC}$: Power consumption of LC.
\item $P_{ONU}$: Power consumption of ONU.
\item $P_{Tx},P_{FH}$: Transmit power and fronthaul link power consumption per radio unit.
\item $P_{CC}^{cool}, P_{EC}^{cool}$: Power consumption of cooling at CC/EC.
\item $P_{CC}^{cache}, P_{EC}^{cache}$: Power consumption of cache processing at CC/EC.
\end{itemize}

\subsubsection{Delay parameters} \label{sec:delay}

\begin{itemize}
\item $D_{prc}(p_i,q_c)$: Delay induced by functions processing given a specific split decision and is calculated as, 
\begin{multline}
D_{prc}(p_i,q_c) = \sum_{i\in[p_i,F_{UP}]}d_{i,prc}^{CC} + \sum_{i\in[0,p_i]}d_{i,prc}^{EC} + \\ \sum_{i\in[q_c,F_{CP}]}d_{i,prc}^{CC} + \sum_{i\in[0,q_c]}d_{i,prc}^{EC}
\end{multline}
%
%
where the first two terms denote the UP processing delay at CC and EC and the last two terms show that of CP at CC and EC. Each delay component is function of the equipment processing speed and required processing \cite{alabbasi2017delay}.
\item $D_{rsf}:$ Delay induced by the number of radio subframes and is calculated as,
\begin{equation}
D_{rsf} = N_{rsf}T_{rsf}
\end{equation}
where $T_{rsf}$  is radio subframe transmission time and  $N_{rsf}$ is the number of radio subframes and is given by,
\begin{equation}
N_{rsf} = \Big[\frac{S_f}{u_{prb}u_{MI}N_{s}}\Big]
\end{equation}
 where $N_{s}$ is the number of symbols per physical resource block (PRB), $u_{prb}$ is the number PRB, and $u_{MI}$ is the modulation index in $bits/prb$.
\item $D_{N_{of}}:$ Delay induced by the number of optical frames.
\begin{equation}
D_{N_{of}} = N_{of}(p_i,q_c)T_{of}
\end{equation}
where, $T_{of}$ is the optical frame time and $N_{of}(p_i,q_c)$  denotes the required number of optical frames to transmit UP/CP data which is given by,  
\begin{equation}
N_{of}(p_i,q_c) = \Big[\frac{V^{cc}(p_i,q_c)N_{rsf}}{S_{of}/|\mathbb{C}|}\Big]
\label{opticalframe_up}
\end{equation}
where $V^{cc}(p_i,q_c)$ is data resulted from CP/UP \cite{virtualization2016functional}. $S_{of}$ is the optical frame size and the denominator of (\ref{opticalframe_up}), shows the amount of bytes that can be used by cell $c$ in an optical frame in the midhaul. Note that we assume that the function processing delay is calculated individually for each radio subframe, then accumulated for all radio subframes.

\item $D_{ONU},D_{LC}:$ Delay induced due to ONU and LC.
\item $D_{opg}:$ Delay incurred due to optical propagation.
\item $D_{mWprg},D_{mWcnv}:$ Delay due to mm-Wave propagation and mm-Wave conversion.
\item $D_{rpg}:$ Delay due to radio propagation and is calculated by dividing the user's distance to the RU by speed of light. 
\item $D_{sw}:$ Delay due to switches.
\item $D_{CC}^{cache},D_{EC}^{cache}:$ Delay incurred due to cache processing at CC and EC.
\end{itemize}
The calculation of delay parameters is explained in \cite{alabbasi2017delay}.

\subsection{Variables}
\begin{itemize}
\item \textbf{$p_i \in [0,F_{UP}]$ :} Integer variable denoting the UP functions split of UE $i$. Larger number of UP are at EC for higher value of $p_i$, hence, if $p_i$ = $F_{UP}$ then all UP functions are distributed, otherwise, if $p_i$ = 0 then all UP functions are centralized.
\item \textbf{$q_c \in [0,F_{CP}]$:} Integer variable denoting the CP functions split of cell $c$. Larger number of CP are at EC for higher value of $q_c$. Hence, if $q_c$ = $F_{CP}$ then all CP functions are distributed, otherwise, if $q_c$ = 0 then all CP functions are centralized.
\item \textbf{$m_i \in D_r$:} Integer variable indexing the DU hosting UPs of UE $i$ at EC $r$. Note that since the association between $i$ and $r$ is fixed, UE $i$ can choose a DU from a given set.
\item \textbf{$n_i \in D_{-1}$:} Integer variable indexing the DU hosting UPs of UE $i$ at CC.
\item \textbf{$x_c \in D_r$:} Integer variable indexing the DU hosting CPs of cell $c$ at EC $r$.
\item \textbf{$y_c \in D_{-1}$:} Integer variable indexing the DU hosting CPs of cell $c$ at CC.
\item \textbf{$w_r$:} Integer variable indexing the wavelength used by EC $r$.
\item \textbf{$l_r$:} Integer variable denoting number of active DUs at EC $r$. $l_r = \sum_{d\in \mathbb{D}_r}\big ( I(x_1=d) \oplus \dots \oplus I(x_{|\mathbb{C}|}=d) \big),$ 
where $\oplus$ shows OR function.

\item \textbf{$l$:} Integer variable denoting number of active DUs at CC. $l = \sum_{d\in \mathbb{D}_{-1}}\big ( I(y_1=d) \oplus \dots \oplus I(y_{|\mathbb{C}_0|}=d) \big),$ where $\oplus$ shows OR function.

\item \textbf{$g$:} Integer variable denoting number of active wavelengths in the midhaul. $g= \sum_{w\in \mathbb{W}}\big ( I(w_1=w) \oplus \dots \oplus I(w_{|\mathbb{R}|}=w) \big), $ where $\oplus$ shows OR function.
\item\textbf{$b_{f,r,{i}}$:} Binary variable denoting if file $f$ is placed at EC $r$ for user $i$.
\item \textbf{$\delta_i$:} Binary variable denoting if user's requested file is at EC and is calculated as $\delta_i =\sum_{r\in\mathbb{R}}\sum_{f\in\mathbb{F}}b_{f,r,i}
$.
\end{itemize}



\subsection{Constraints}
Constraint (\ref{cons 1}) ensures that the functional split of baseband processing can at most occur once at CP or UP.
\begin{equation}
I(p_i<F_{UP})+I(q_c<F_{CP}) = 1, \qquad \forall i \in \mathbb{I}_c,\forall c \in \mathbb{C}_o.
\label{cons 1}
\end{equation}
Constraint (\ref{cons 2}) ensures that, if the UP of UE $i$ is split, then all the UP and CP below the split point must be placed at the EC processed by the same DU.
\begin{equation}
I(p_i<F_{UP}) \Rightarrow (m_i = x_c), \qquad \forall i \in \mathbb{I}_c,\forall c \in \mathbb{C}_o.
\label{cons 2}
\end{equation}
Constraint (\ref{cons 3}) ensures that, if CP of cell $c$ is split, then all the CP and UP above the split point must be placed at CC processed by the same DU. 
\begin{equation}
I(q_c<F_{CP}) \Rightarrow (n_i = y_c), \qquad \forall i \in \mathbb{I}_c,\forall c \in \mathbb{C}_o.
\label{cons 3}
\end{equation}
Constraint (\ref{cons 4}) ensures that, the total number of CP processed by a DU $d$ at EC must not exceed the maximum capacity of the number of CP that can be hosted by a DU at EC. 
\begin{equation}
\sum_{c\in \mathbb{C}_r}H_{CP}^{EC}(q_c).I(x_c=d) \leq L_{CP}^{EC},  \forall r \in \mathbb{R},\forall d \in \mathbb{D}_r.
\label{cons 4}
\end{equation}
Constraint (\ref{cons 5}) ensures that, the total number of CP functions processed by a DU $d$ at CC must not exceed the maximum capacity of the number of CP functions that can be hosted by a DU at CC. 
\begin{equation}
\sum_{c\in \mathbb{C}_o}H_{CP}^{CC}(q_c).I(y_c=d) \leq L_{CP}^{CC}, \qquad \forall d \in \mathbb{D}_{-1}.
\label{cons 5}
\end{equation}
Constraint (\ref{cons 6}) ensures that the number of UPs that are accommodated by a DU $d$ at EC $r$ cannot exceed this EC-DU’s UP capacity.
\begin{equation}
\sum_{c\in \mathbb{C}_r}\sum_{i\in \mathbb{I}_c}H_{UP}^{EC}(p_i).I(m_i=d) \leq L_{UP}^{EC}, \\ \forall r \in \mathbb{R},\forall d \in \mathbb{D}_r.
\label{cons 6}
\end{equation}
Constraint (\ref{cons 7}) ensures that the number of UPs that are accommodated by a DU $d$ at CC cannot exceed this CC-DU’s UP capacity.
\begin{equation}
\sum_{i\in \mathbb{I}_o}H_{UP}^{CC}(p_i).I(n_i=d) \leq L_{UP}^{CC}, \qquad \forall d \in \mathbb{D}_{-1}.
\label{cons 7}
\end{equation}
Constraint (\ref{cons 8}) ensures that the total occupied midhaul bandwidth in a wavelength cannot exceed the wavelength’s capacity, i.e., $K$.
\begin{equation}
\sum_{r \in \mathbb{R}}I(w_r=w).\sum_{c\in \mathbb{C}_r}\Big(G_c(q_c)+\sum_{i\in \mathbb{I}_c}J_i(p_i)\Big) \leq K, \\ \forall w \in \mathbb{W}.
\label{cons 8}
\end{equation}
Constraint (\ref{cons 9}) and (\ref{cons 10}) ensures that, if the function processing is at CC then content cannot be placed at local EC. If the content is optimally placed at local EC, then function processing must be at local EC.
\begin{equation}
p_i - F_{UP} \leq M(1 - b_{{f,r},i}), \,\,\,\forall i\in \mathbb{I}_c,\forall r\in \mathbb{R},\forall f\in \mathbb{F}.
\label{cons 9}
\end{equation}
\begin{equation}
p_i - F_{UP} \geq -M(1 - b_{{f,r},i}), \,\,\,\forall i\in \mathbb{I}_c,\forall r\in \mathbb{R},\forall f\in \mathbb{F}.
\label{cons 10}
\end{equation}
where $M$ denotes a big number for the big M method optimization. Constraint (\ref{cons 11}) ensures that the capacity of cache at EC $r$ cannot exceed the maximum capacity of the EC's cache
\begin{equation}
X_r \leq C_{EC}, \qquad \forall b\in \mathbb{B}.
\label{cons 11}
\end{equation}
where
\begin{equation}
X_r = \sum_{f \in \mathbb{F}_r}I(b_{{f,r},i}).{S_f}, \qquad \forall r \in \mathbb{R}.
\end{equation}
Constraint (\ref{cons 12}) ensures that the total delay of a user i should be less than the delay threshold 
\begin{equation}
D_{i} \leq d_i.
\label{cons 12}
\end{equation}
where $D_{i}$ for user i is given by,
\begin{multline}
D_{i} = D_{prc}(p_i,q_c) + D_{rsf} + D_{N_{of}} + D_{ONU} + D_{LC} + \\ D_{opg} + D_{mWprg} + D_{mWcnv} + D_{rpg} + D_{sw} + \delta_i D_{EC}^{cache} + \\ (1-\delta_i)D_{CC}^{cache}.
\end{multline}
 All delay parameters are explained in Section \ref{sec:delay}. 
We solved the proposed problem optimally using IBM ILOG CP solver. This solver can solve the problem with the defined optimal gap. We can solve the problem optimally provided that the optimality gap is set to 0 \cite{manual2015ibm}.



\section{Simulation Results}\label{Sec:Sim}

\begin{table}\footnotesize \centering
\caption{Simulation parameters}
\label{my-label}
\begin{tabular}{|p{0.47\columnwidth}|p{0.44\columnwidth}|}
\hline
\textbf{PARAMETERS} & \textbf{VALUES} \\ \hline
Topology & \begin{tabular}[c]{@{}l@{}}1 CC, 4 ECs, 5 RU per EC, \\ each RU serve up to 5 users\end{tabular} \\ \hline
Configuration of RU & 20 MHz, 2*2 MIMO, 64 QAM \\ \hline
Capacity of DU at EC & 3 CPs / 15 UPs \\ \hline
Capacity of DU at CC & 37 CPs / 135 UPs \\ \hline
Size of the requested file & 20 MB \\ \hline
Capacity limit of midhaul link & 26000 Mbps \\ \hline
Number of CP/UP ($F_{UP}$ , $F_{CP}$) & $F_{UP}$ = 3, $F_{CP}$ = 3 \\ \hline
Power of DU at EC/CC & 50 W / 100 W \\ \hline
LC power + ONU power & 20 W, 5 W \\ \hline
\begin{tabular}[c]{@{}l@{}}Radio access + Fronthaul \\ link power consumption\end{tabular} & 20 W + 40 W \\ \hline
Power  caching  at EC/CC & 30 W / 20 W \\ \hline
Optical frame size, $S_{of}$ & 38880 bytes \cite{sectorseries} \\ \hline
Delay of (ONU,LC) & (7.5,1.5)*1e-6 sec    \\ \hline
Cache processing delay at EC/CC & (25.0/20)*1e-3 sec \\ \hline
Delay of optical transmission & 0.4 * 1e-3 sec \\ \hline
(Ethernet,Optical) switching delay & (5.2 * 1e-3,2.5) * 1e-3 sec \\ \hline
Mm-wave conversion delay & 30 * 1e-6 sec \\ \hline
\end{tabular}
\end{table}

\useunder{\uline}{\ul}{}




In this section, the performance of joint functional split and content placement (FSCP) in H-CRAN is evaluated. We consider three key performance metrics, namely 1) network power consumption, 2) content access delay, and 3) hit rate, where the hit rate is defined as the ratio of served users with satisfied demands, e.g., delay, to the total number of users. We assume maximum number of 95 users are connected to the RUs of the corresponding cell. 
  Both the EC and CC are equipped with the DUs which have different maximum capacity of hosting CPs and UPs given in Table \ref{my-label}. 
 We assume that users are distributed uniformly in the cell coverage area of $250m$ and one resource block is assigned to each user. In our system, each user requests randomly with a uniform distribution, a content of size $S_f$ from a set $\mathbb{F}$, within a delay threshold $d_i$.  All the simulation parameters are summarized in  Table \ref{my-label}. FSCP is compared with 2 reference cases where files are cached at EC (1) and CC (2).

\begin{figure}[!t]
\centering
    \begin{subfigure}[t]{0.44\textwidth}
  \includegraphics[width=1\textwidth]{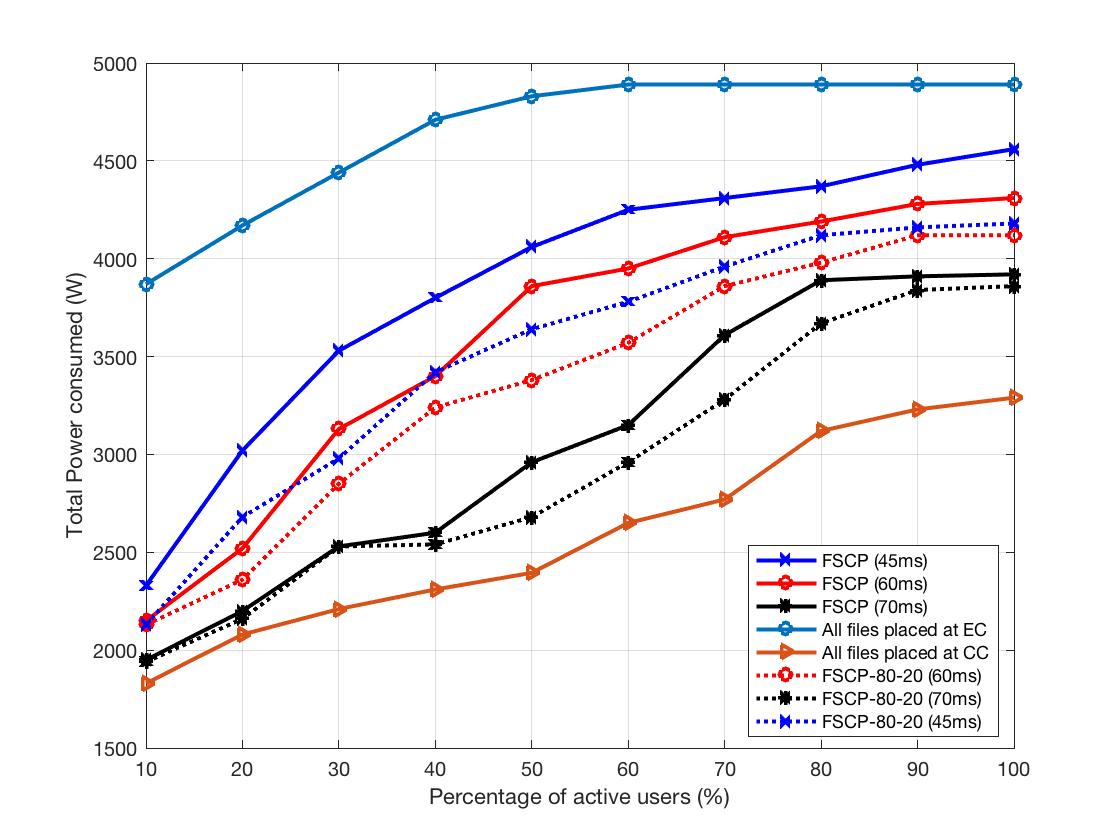}
  \caption{Power consumption.}
  \label{fig:powerconsumption}
 \end{subfigure}
    \begin{subfigure}[t]{0.44\textwidth}
  \includegraphics[width=1\textwidth]{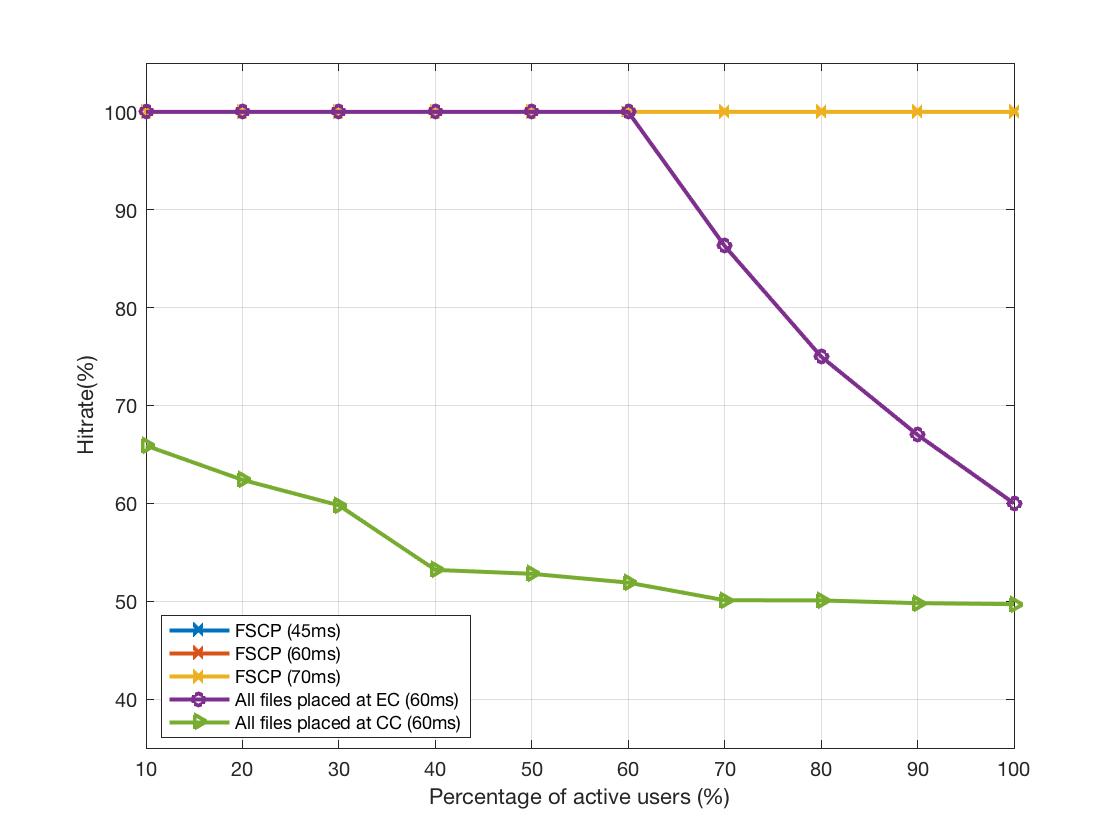}
  \caption{Hit rate.}
  \label{fig:hitrate_power}
\end{subfigure}
  \caption{Power/hit rate against number of active users.}
  \label{fig:power_hitrate_power}
\end{figure}


Fig. \ref{fig:powerconsumption} illustrates the impact of the number of active users on total power consumption. There is high possibility that multiple users request for the same content in a cell. To cover this request pattern, the user request was modeled such that 80 percentage of the users of the same EC request for the 20 percent of the total contents (FSCP-80-20). Placing all the contents at EC restricts the centralization of the functions at EC according to constraint (9), thus it consumes more power due to low computational capacity of the DUs at EC. In this case, after a point, we can see a saturation in power consumption. This saturation is because of the capacity of EC which is fully utilized and has no room to accommodate users anymore. When all contents are placed at CC, the least power is consumed, however, in this case some of users are dropped due to the latency constraint. FSCP power consumption goes down as the mean delay threshold increases, i.e., from 45msec to 70msec.
When the contents can be stored as in FSCP-80-20 ,up to 8.33$\%$ more power consumption can be saved compared to FSCP.
It is worth mentioning that one source of power saving is due to centralizing the processing at the CC and taking advantage multiplexing gain among all cells. Another source of power saving is due to  turning off the unused components in the network such as DUs, LCs, ONUs, and etc. Finally, if contents are stored at edge and are requested by multiple users, cell processing functions are done only once for those users and hence less power is consumed. In Fig. \ref{fig:hitrate_power}, the system is evaluated based on the hit rate ratio. FSCP attains full success hit rate. At full load and in all cache at CC scenario, half of users are served,  while $60\%$ are served in EC scenario. The former is due to delay constraints and the latter is because of storage capacity limitation in the EC. There is a breaking point in EC scenario where beyond that point the EC can not serve more users. This point corresponds to the saturation point in Fig. \ref{fig:powerconsumption}.





\begin{figure}[t]
  \begin{center}
  \includegraphics[width=.9\columnwidth]{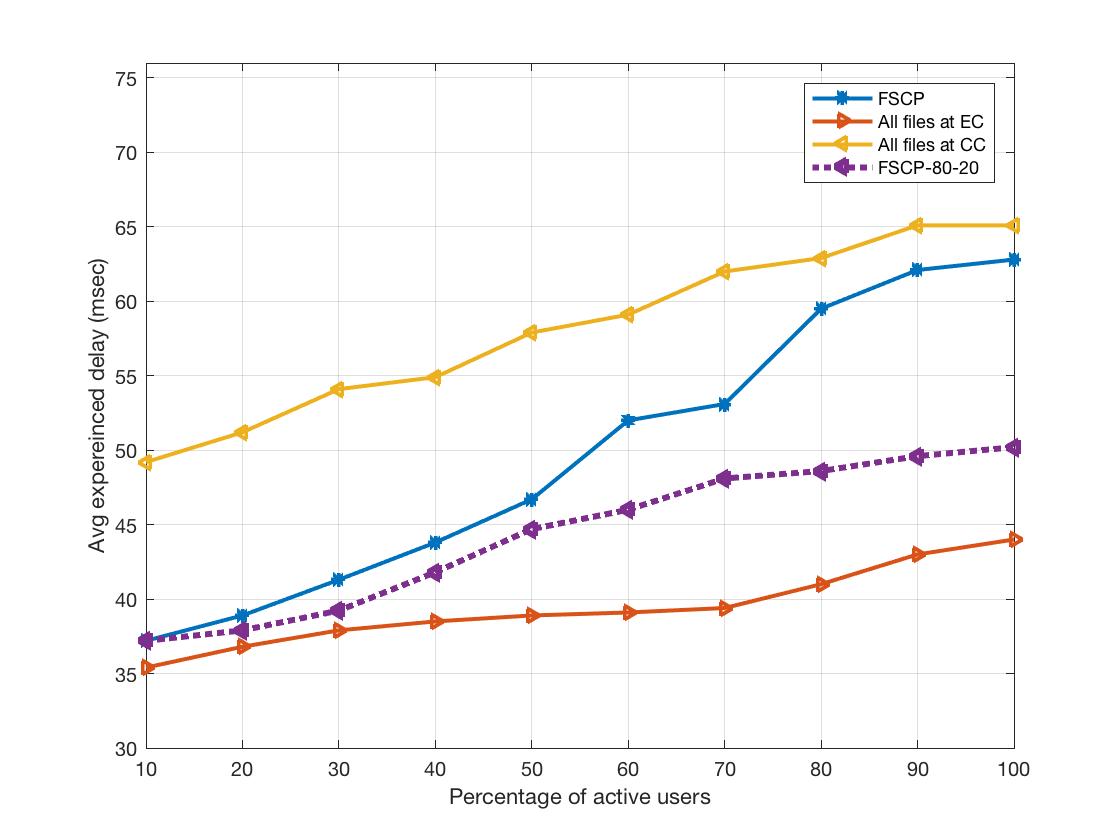}
  \end{center}\footnotesize
  \caption{Average delay of the users vs percentage of active users}
  \label{fig:avg_delay}
\end{figure}

In Fig. \ref{fig:avg_delay}, we have depicted the average delay experienced by the users. 
Fetching the content from the CC experiences more delay than other cases due to the higher distance between the CC and the users. FSCP-80-20 has up to 20$\%$ lower delay compared to FSCP due to the content sharing. Comparing this trend with that of Fig. \ref{fig:powerconsumption}, one can see the trade off between the experienced delay and the the total power consumption. 


\begin{figure}[t]
  \begin{center}
  \includegraphics[width=.9\columnwidth]{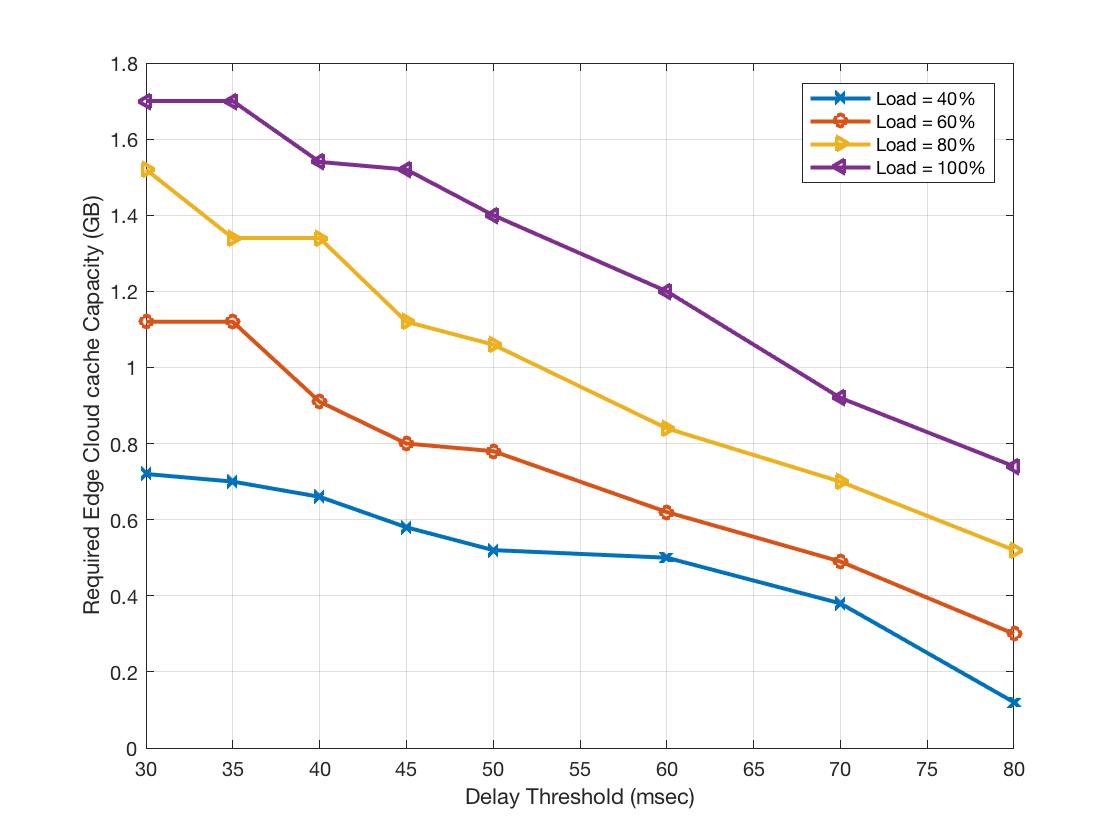}
  \end{center}\footnotesize
  \caption{Impact of delay threshold on the required EC capacity for FSCP}
  \label{fig:dlythesh_cap}
\end{figure}
Fig. \ref{fig:dlythesh_cap} shows the impact of the delay threshold over the total required edge capacity for content storage for various traffic load. By increasing the delay threshold, contents are likely to be stored at CC to benefit from the efficient computation capacity of DUs at CC. This means, we only cache the contents at EC if it is necessary and thus we require less capacity at EC.
\begin{figure}[t]
  \begin{center}
  \includegraphics[width=.9\columnwidth]{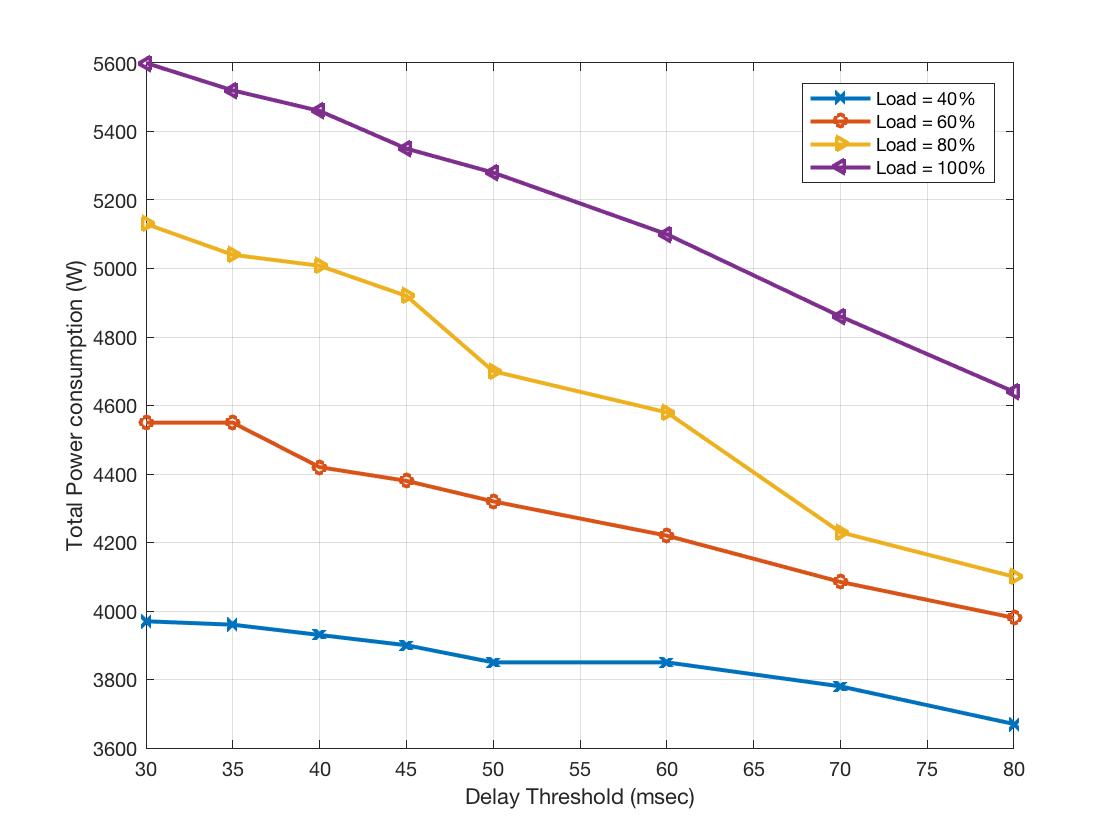}
  \end{center}\footnotesize
  \caption{Delay threshold impact  on the total power consumption for FSCP}
  \label{fig:dlythresh_power}
\end{figure}
Fig. \ref{fig:dlythresh_power} shows the impact of the delay threshold over total network power consumption. Increasing the delay threshold allows the users to access the content from the CC, thus allows the flexibility of processing the functions at CC, subsequently decreases the power consumption.

\section{Conclusion} \label{Sec:con}
In this paper, we considered the joint problem of content placement and functional split to minimize power consumption of hybrid CRAN architecture. We proposed an optimization framework to jointly perform functional split and content caching  by developing a constraint programming model. Users' QoS, defined as the delay requirement, is also considered. Numerical results illustrated that compared to CRAN architecture, in low load scenario, the experienced delay of Hybrid CRAN is 10\% lower at cost of 20\% more power consumption knowing that QoS of all users are satisfied. Compared to edge caching, in low load scenario, about 35\% power saving is attained at cost of 10\% more experienced delay. In summary, utilizing the  content caching at edge cloud together with functional splitting is beneficial in terms of  content access delay but at cost of power consumption. Since the formulated problem is NP-hard meaning, one research direction is to use heuristic algorithm or artificial intelligence to solve the problem. Moreover, considering the content popularity for each content would provide more realistic approach towards the content placement problem.


\bibliographystyle{IEEEtran}
\bibliography{bibs}
\end{document}